\documentstyle[12pt]{article}
\begin{document}

\begin{center} Gonihedric String and Asymptotic Freedom
\end{center}
\begin{center}G.K.Savvidy
\end{center}
\begin{center}Institut f\"ur Theoretische Physik der J.W.Goethe-
Universit\"at, D-6000 Frankfurt am Main 11, Germany
\end{center}
\begin{center}and
\end{center}
\begin{center}Yerevan Physics Institute, Yerevan 375036, Armenia
\end{center}
\begin{center}K.G.Savvidy
\end{center}
\begin{center}Moscow Institut of Physics and Technology,
141700 Moscow, Russia
\end{center}
\begin{center}Abstract
\end{center}

   Few natural basic principles allow to extend  Feynman
integral over the paths to an integral over the surfaces so,
that they coincide at long time scale, that is when the surface
degenerates into a single particle world line. In the first
approximation the loop Green functions have perimeter behavior.
That corresponds to the case when free quarks interact through
one gluon exchange. Quantum fluctuations of the surface generate
the area low. Thus in this string theory asymptotic freedom and
confinement can coexist.
\newpage

   There are few important facts constituting our knowledge of strong
interaction. The hadrons are made up of quarks and they are free to
propagate together through the space-time.However it is impossible to
separate them without creating  new hadrons and, contrary to that
picture in deep-inelastic processes, over short time and short
distances they behave like free pointlike particles
\cite{gellmann,feynman,bjorken,paschos,gross,politzer,wilson}.

   The hadrons, propagating as a whole through the space-time,
swept out
the world line, narrow strip, and Feynman path integral perfectly
describes this propagation. To describe the fine
structure of hadrons
one should take into account an internal motion of the quarks.
Perhaps the gauge field, keeping quarks together, compresses into
flux line and the propagation of hadron through the space-time
forms the world surface rather than the line
\cite{wilson,hooft,schwinger,kogut,casher}.

   Thus, to describe the propagation of the real hadrons with
internal structure one should extend Feynman integral over the
paths to integral over the surfaces in a such way that they
will ${\it coincide}$ at long time scale, that is in the cases, when
the surface degenerates into a single particle world line
\cite{savvidy}.

   To incorporate these properties into the theory one should
require that i) long narrow strips of the world surface must have
the amplitude proportional to the average length of the strip
\cite{savvidy}
ii) two surfaces distinguished by a small deformation of the
shape should have close amplitudes \cite{ambar,savvidy}
and iii) "fat" surfaces
should appear with small amplitude\cite{wilson}.

   Feynman path integral is defined by summing over all
piecewise linear random walks from point $X$ to $Y$ with
transition amplitude proportional to the length of the
path $A(L)$ \cite{feynman1}
$$K(X,Y) = \sum_{paths} e5{-A(L)} \approx \int \frac{e5{ip(X-Y)}}
{p5{2} + m5{2}} \,dp \eqno(1)$$

   While the surface is identified with a piecewise flat triangles
(polygons) glued together through their sides and as we stressed
before the amplitude should be proportional to the length i).
Thus it must be proportional to the linear combination of the
lengths of all surface edges

$$A(M) = \sum_{<i,j>} \Theta_{i,j} \cdot \vert X_{i} -X_{j} \vert$$
where $X_{i}$ denotes the vertex coordinate, summation is over all
edges $<X_{i},X_{j}>$ (i and j are the nearest neighbors) and
$\Theta_{i,j}$ is unknown factor, which can be defined by use of
the requirement ii). Indeed, if we impose a new vertex $X$ on a given
flat triangle $<X_{1},X_{2},X_{3}>$, then for that new surface
we will get an extra contribution $\sum_{i} \Theta_{i}
\cdot \vert X -X_{i} \vert$ to $A(M)$ and we will get more extra
terms imposing
a new vertexes, despite the fact that the surface does not
change visible. To exclude such tape of contributions we should choose
unknown factor $\Theta_{i,j}$ such that it will vanish
in flat cases. This can be done by use of the edge angles, therefore
\cite{ambar,savvidy}
$$A(M) = \frac{1}{2} \sum_{<i,j>} \vert X_{i} -X_{j} \vert
\Theta(\alpha_{i,j}) \eqno(2)$$
where

$$\Theta(\pi) = 0, \eqno(3)$$
$$\Theta(2\pi-\alpha) = \Theta(\alpha), \eqno(4)$$
$$\Theta(\alpha) \geq  0 \eqno(5)$$
and $\alpha_{i,j}$ is the angle between the neighbor triangles in
$R5{d}$ having a common edge $<X_{i},X_{j}>$. First condition (3)
guarantees that property ii) is fulfilled, the second (4), that the
action (2) is local, that is one can measure the angle $\alpha_{i,j}$
irrespective to the side of the surface and the third one (5)
that the action is positive. These conditions do not define
the weight factor $\Theta(\alpha)$ completely, continuous "number"
of possibilities still remain. Thus string should be further
tuned to convenient scaling behavior by the appropriate selection
of the factor $\Theta(\alpha)$\cite{savvidy} .

  For that let us consider a surface with the boundaries created
by external sources or by virtual quarks. It is reasonable to take
$\alpha_{i,j}$ on the boundary edges equal to zero or to $2\pi$.
Then from (2) it follows that the boundary part of the action
is simply proportional to the full length of the boundary

$$A_{\it boundary} = \frac{1}{2} \cdot L \cdot \Theta(0) \eqno(6)$$
In (6) the total length of the boundary $L$ is multiplied
by the factor
$\Theta(0)$, thus $\Theta(0)$ plays the role of the hopping
parameter $K$ of the quarks

$$\Theta(0) = 2 \ln \frac{1}{K}  \eqno(7)$$
and if quarks are infinitely heavy, then

$$\Theta(0) \rightarrow \infty \eqno(8)$$
and is of order one for massless quarks.

   Now let us consider vacuum expectation values of two currents
$<J(X),J(Y)>$ built from quark fields. In this approach it should
be represented by the sums over all quark paths and sums over all
surfaces connecting these paths. All possible virtual quark loops
must be summed over too, that is over surfaces  with holes (rimmed
by quark loops)

\vspace{1cm}

$<J(X),J(Y)> = \sum_{paths} \sum_{surfaces}$

\vspace{1cm}
As we know (6), the amplitude  associated with quark paths is
equal to

$$e5{-\frac{1}{2} \cdot P \cdot \Theta(0)} \eqno(9)$$
where $P$ is the total length of the quark lines, including the
internal quark loops. In the limit of infinitely heavy quarks (8),
$\Theta(0)
\rightarrow \infty $, the virtual quark loops are suppressed (9),
thus only valence quark and antiquark paths connecting $X$ and
$Y$ will remain. Before that quark paths averaging the Green function
can be represented in the form

$$K(X,Y) = e5{-\frac{1}{2} \cdot P \cdot \Theta(0)}
\cdot \sum_{T} \int e5{-A_{inside}(M)} \cdot \prod_{i \in
T / \partial T} dX_{i} \eqno(10) $$
where $T$ denotes the set of triangulations and integration
is over all vertexes inside the loop with fixed boundary.
The first term in (10)
represents the perimeter low and is expected when free quarks
interact through one gluon exchange,as it takes place in
heavy meson systems or in $e5{+}e5{-}$ annihilation
\cite{feynman,bjorken,paschos,gross,politzer,wilson}.
It is very difficult to satisfy this important property
of strong interaction in string theories with area action
\cite{nambu,nielsen,susskind} or with gaussian action
\cite{durhuus}. The second
term in (10) is associated with quantum fluctuations of the
surface and will produce quantum corrections to a perimeter
low (9). Below we will see that these fluctuations generate
confinement \cite{savvidy}, that is the area low iii).
To simplify the proof
let us consider the case of infinitely heavy quarks (8). Indeed,
when $\Theta(0) \rightarrow \infty$, the crumpled surfaces are
strongly suppressed, because for them $\alpha_{i;j}$ is
close to zero and $\Theta(\alpha)$ is actually very large.
Thus,surface fluctuates near the angles close to $\alpha \approx
\pi$, that is it is flat and therefore $\Theta(\alpha) \approx
0$. If $X_{\perp}$ is the amplitude of transverse fluctuations,
then angles fluctuate in the region of order $X_{\perp}/(P/
\sqrt{T})$, where T is the number of vertexes. The action is
equal to

$$A_{inside}(M) \approx T \cdot \frac{P}{\sqrt{T}} \cdot
\left( \frac{X_{\perp} \sqrt{T}}{P} \right)5{\varsigma},$$
where we use the following parameterization of $\Theta(\alpha)$
near $\alpha \approx \pi$

$$\Theta(\alpha) = \vert \pi - \alpha \vert 5{\varsigma}.$$
Therefore

$$\sum_{T} \int e5{- \beta A_{inside}(M)} \cdot \prod_{i \in
T/  \partial T} dX_{i} \approx
\sum_{T} \left( \frac{P5{2}}{T} X_{\perp}5{d-2} \right)5{T}
e5{-\beta P\sqrt{T} \left( \frac{X_{\perp} \sqrt{T}}{P}
\right)5{\varsigma}} \eqno(11)$$
and expressing $\varsigma$ in the form

$$\varsigma = \frac{d-2}{d+2\delta} \eqno(12)$$
for the transverse fluctuation we will get

$$<X_{\perp}>  \approx \left( \frac{d+2\delta}{\beta} \right)5
{\frac{d+2\delta}{d-2}}\cdot \left( \frac{\sqrt{T}}{P} \right)5
{\frac{2+2 \delta }{d-2}} \eqno(13)$$
Substituting (13) into (11) yield

$$\sum_{T} e5{-\delta \cdot T \cdot \ln(P5{2}/T) - (d+2\delta)
\cdot T \cdot (1-ln(\frac{d+2\delta}{\beta}))}\eqno(14)$$
and only for $\delta = 0$ we will have convergence, hence

$$\sum_{T} e5{-T \cdot d (1- \ln(d/ \beta))} \approx
e5{\sigma (\beta) \cdot S} \eqno(15)$$
where $S$ is the, area and the string tension $\sigma(\beta)$
is equal to

$$\sigma(\beta) \approx \frac{d}{a5{2}} (1-\ln(d/ \beta))
\approx \frac{\beta - \beta_{c}}{a5{2}} \eqno(16)$$
So string tension scales and the exponent $\nu$ is equal to

$$\nu = \frac{1}{d_{H}} = \frac{1}{2},\eqno(17)$$
where
$$a= (\beta - \beta_{c})5
{\frac{1}{2}}. $$
This value coincide with the one for the random walk integral
(1) and is expected, because both theories are close in nature
i). Our choice of $\delta$ corresponds to the full compensation
of the longitudinal and transverse fluctuations

$$\frac{P5{2}}{T} X_{\perp}5{d-2} = 1 $$

The physical reason, why this model generates confinement,
is very simple. As we stressed in ii) the vertexes which lie
on the flat minimal surface do not contribute into the action,
that is they "prefer" to fluctuate near flat minimal surface
occupying
the area $P5{2}/T$ each. Now fluctuations in the transverse
direction are regulated by "transverse" action in (11) and they are
small if $\delta $ in (12) is sufficiently large. Thus the
Wilson integral is proportional to the volume of the narrow
layer around the minimal surface. Confinement
appears as a natural consequence of the basic principles ii).

   Suppose now, that quarks are rather light, and thus $\Theta$
is not very large. Then fluctuations with acute edge angles,
$\alpha_{i,j} \approx 0$, will have large amplitude and crumpled
surfaces are no longer suppressed. Now let us consider a new
surface which has a cuts along these acute edges. These two surfaces
have almost the same amplitude. This fact is a direct consequence
of the continuity property ii), the factor $\Theta(\alpha)$ is
unique for all surface and therefore, there is no big difference
between contributions from boundary edges, where $\alpha_{i,j} =
0$ or from the acute edges, where $\alpha_{i,j} \approx
0$. This means that virtual quark loops can now easily
be created
and we know that this take place when a surface crumples
producing edges with acute angles.

   Together with our previous result (15), that the "fat"
surfaces are strongly suppressed, quark and antiquarks paths
are unlikely to be well separated, this drives
us to the conclusion
that surface should easily tear into pieces along the acute
edges, producing long narrow strips, the mesons, which can
propagate
through the space-time with the amplitude (1) as it follows
from i) and (2). This is in accord with Wilson picture
\cite{wilson},
the binding mechanism is soft, because
we get perimeter low for the cases of nearby
quark-antiquark pairs and exponential
damping is associated with large size loops having large
area. This picture is very close to what
we are expecting out of QCD.

   We don't know how to introduce baryons into the theory
in a unique way, but if one consider the baryons as a
surface having three strips glued together through
one common boundary and therefore forming the letter
$Y$, one should only define the extra contribution
coming from common side-edges. By use of the  requirement ii)
it is natural to take these common edges with the weight

$$\sum_{common edges} \vert X_{i} - X_{j} \vert \cdot
(\Theta(\alpha_{i,j}) + \Theta(\beta_{i,j}) +
\Theta(\gamma_{i,j})) \eqno(18)$$
where $\alpha , \beta ,\gamma$ are the angles between
strips. It is easy to consider the cases, when this surface
describe the decay of the baryon.

   Few remarks are in order. First of all, when $d \rightarrow
\infty$ in (12), then $\varsigma = 1$,
$\Theta(\alpha) = \vert \pi - \alpha
\vert$ and the action (2) coincides with the modified Steiner
functional, which has different representations in term
of geometrical invariants \cite{savvidy}.
One of them, for the convex surfaces
is represented through the mean size of the surface

$$M = \frac{1}{2} \int L_{g} dg , \eqno(19) $$
where $L_{g}$ is the orthogonal projection of the surface to
the line $g$ and integration is over all lines crossing the
origin. Minkowski representation (19) tells how to "measure"
the surface in terms of "meters" rather than in "square meters".
Because $L_{g}$ is always bigger than the projection of the
diameter $\Delta$ of the surface $L_{g} \geq
\Delta \cdot \vert \sin(\varphi) \vert$
($\Delta$ is the maximal
distance between two points on the surface),
we have the bound
$M \geq  \Delta$ and therefore the upper bound for
the partition function\cite{savvidy}

$$Z_{T} \leq (Const)5{T}(dT)! \eqno(20)$$
where $Z_{T}$ is the contribution of the given triangulation
$T$ to the full partition function $Z=\sum Z_{T}$
and as we see it is finite. It is also
true, that $\Delta \geq L_{g}$, therefore we have the same
expression for the lower bound and thus the answer (20) is exact
for the modified Steiner functional.
For nonconvex surfaces there is an extra contribution from
"inside" edges and we have instead of (19)

$$A(M) \geq \frac{1}{2} \int L_{g} dg,$$
thus the upper bound (20) remains correct in all cases, that is
$Z_{T}$ is finite. Now, to get a convergence of the full
partition function we should take  $\varsigma < 1$.
Indeed, for $\varsigma < 1$ the action
increases much faster for crumpled surfaces, compared with
the case
when $\varsigma = 1$ and, as it is seen from
(14) and (10), we get the convergence.

The other expression
for the Steiner functional is realized
through the extrinsic curvature
or through the mean value of the length of the orthogonal
projection of the surface on the two-dimensional plane. But
all these expressions have limiting validity. The expression
(2) is more reliable, because it is well defined for
extremely large class of surfaces with arbitrary topology
and in any dimension. It coincides with other expressions in
special cases.

The second remark is connected with our derivation of the
area low.
One can see, that this derivation takes into account
only high frequency modes of the surface fluctuations and one
should  therefore account low frequencies too. Indeed
they produce
corrections, but do not change the result.

And the last one is connected with the string theories, which
are based on the area action \cite{jonsson,sukiasian}.
The main problem here is connected
with the short wave-length fluctuations. It is very hard
to avoid instantaneous excitement of these modes
in deep inelastic scattering.

As we have seen, asymptotic freedom and confinement can
coexist in gonihedric string theory.

We are grateful to J.Ambjorn, E.Floratos, R.Flume,
H.Fritsche, D.Gross, C.Kounnas, P.Minkowski, H.Nielsen
and E.Paschos for fruitful discussions. One of the author
(G,K.S.) is thankful to W.Greiner for discussions and warm
hospitality in Frankfurt University.

This work was done under financial support of Alexander von
Humboldt Foundation.

\end{document}